\documentclass[12pt]{iopart}

\usepackage{graphicx}
\usepackage{epstopdf}

\begin{document}

\title[Phase transitions in $sdg$-IBM]
{Phase transitions in the $sdg$ interacting boson model}
\author{
P~Van~Isacker\dag,
A~Bouldjedri\ddag\
and S~Zerguine\ddag}

\address{\dag\
Grand Acc\'el\'erateur National d'Ions Lourds,
CEA/DSM--CNRS/IN2P3, BP~55027,
F-14076 Caen Cedex 5, France}

\address{\ddag\
Department of Physics, PRIMALAB Laboratory,
University of Batna, Avenue Boukhelouf M El Hadi,
05000 Batna, Algeria}

\begin{abstract}
A geometric analysis of the $sdg$ interacting boson model is performed.
A coherent-state is used in terms of three types of deformation:
axial quadrupole ($\beta_2$),
axial hexadecapole ($\beta_4$)
and triaxial ($\gamma_2$).
The phase-transitional structure is established
for a schematic $sdg$ hamiltonian
which is intermediate between four dynamical symmetries of U(15),
namely the spherical ${\rm U}(5)\otimes{\rm U}(9)$,
the (prolate and oblate) deformed ${\rm SU}_\pm(3)$ 
and the $\gamma_2$-soft SO(15) limits.
For realistic choices of the hamiltonian parameters
the resulting phase diagram has properties
close to what is obtained in the $sd$ version of the model
and, in particular, no transition towards a stable triaxial shape is found.
\end{abstract}

\section{Introduction}
\label{s_intro}
Phase transitions are well-known phenomena.
They are said to occur if an order parameter of the system
experiences, as function of a control parameter, a rapid change at a particular point.
A phase transition may be of first or second order,
depending on the discontinuity of the first- or second-order derivative
of the order parameter as a function of the control parameter (Ehrenfest classification).
Phase transitions have been observed in countless classical and quantal systems,
usually composed of very large numbers of particles
and hence allowing a statistical treatment.
The atomic nucleus, however, consists of less than 300 particles (nucleons)
and finite-system effects may therefore play a role.

The concept of phase transitions is used in nuclear physics
either to describe the behaviour of hot nuclei resulting from heavy-ion collisions
or in the context of so-called `quantum' phase transitions,
also called shape phase transitions.
The latter is the subject this paper
and relies on a method proposed by Gilmore in 1979~\cite{Gilmore79}.
Its central idea is that any quantum-mechanical hamiltonian
that can be written in terms of the generators of a compact Lie algebra,
has an unambiguously defined classical analogue
with a geometric interpretation in terms of shape variables
of which the phase-transitional behaviour can be studied.
A few years before Gilmore's proposal, in the middle of the 1970s,
Arima and Iachello~\cite{Arima75,Arima76,Arima78,Arima79} had proposed
a novel way to describe the nucleus with a set of $s$ and $d$ bosons,
called the interacting boson model (IBM)~\cite{Iachello87},
which precisely has the property
that it can be cast in terms of generators of the unitary Lie algebra U(6)
to which Gilmore's technique can be applied.
This approach was advocated by Dieperink {\it et al.}~\cite{Dieperink80}
and lead to a geometric interpretation of the IBM~\cite{Ginocchio80,Bohr80}
and to a connection with the Bohr--Mottelson description of the nucleus~\cite{BM75}.
The problem of phase transitions in the IBM
was subsequently addressed by Feng {\it et al.}~\cite{Feng81}
but was solved in its full generality only many years later
by L\'opez-Moreno and Casta\~nos~\cite{Lopez96}.

Although the early work in the 1980s
determined the most essential phase-transitional properties of the IBM,
many additional related features of the model
have been studied over the years~\cite{Iachello98,Casten99,Cejnar00}.
This area of research received renewed impetus
following Iachello's suggestion of
critical-point symmetries E(5) and X(5)~\cite{Iachello00,Iachello01},
nuclear examples of which
were proposed by Casten and Zamfir~\cite{Casten00,Casten01}.
The central idea of Iachello's method is to provide a description
of nuclei at the phase-transitional point
through the (approximate) analytic solution of a geometric Bohr hamiltonian.
This rekindled interest in the phase-transitional behaviour of algebraic models
from various angles:
the correspondence between the E(5) solution of the Bohr hamiltonian
and the U(5)--SO(6) transition in the IBM
was studied in detail~\cite{Frank02,Arias03,Garcia03,Leviatan03,Garcia05},
the existence of an additional prolate--oblate transition
was recognized~\cite{Jolie01,Jolie03},
a connection with Landau theory of phase transitions
was established~\cite{Jolie02,Cejnar03b,Iachello04},
the influence of angular momentum
was investigated~\cite{Cejnar02,Cejnar03,Cejnar04},
the connection with shape coexistence was revisited~\cite{Heyde04},
the algebraic framework of quasi-dynamical symmetries was established
to deal with transitions between two phases~\cite{Rowe04,Rowe04b,Rowe04c,Turner05,Rosensteel05},
phase transitions of excited states were considered~\cite{Caprio08}.
In addition, several extensions of the IBM were re-examined from this angle
such as the neutron--proton version of the model~\cite{Caprio04}
or its configuration-mixed version~\cite{Frank06}.

In this paper we report on a phase-transitional study
of another extension of the IBM, namely the \mbox{$sdg$-IBM},
which includes a $g$ boson
to account for hexadecapole deformations of the nucleus.
We give a brief review of the \mbox{$sdg$-IBM} in section~\ref{s_hamsdg}
and discuss its classical limit in section~\ref{s_clas}.
Since a general $sdg$ hamiltonian has too many interaction parameters,
a simpler parametrization is proposed in section~\ref{s_cast}.
The method for establishing the phase diagram
is explained in section~\ref{s_meth}
for the specific case of the \mbox{$sdg$-IBM}
and then applied to obtain partial and complete diagrams
in sections~\ref{s_part} and~\ref{s_comp}, respectively.
The conclusions of this work are summarized in section~\ref{s_conc}.

\section{The general hamiltonian of \mbox{$sdg$-IBM}}
\label{s_hamsdg}
We refer the reader to the paper of Devi and Kota~\cite{Devi92}
for an excellent review of studies with the \mbox{$sdg$-IBM}.
We limit ourselves here to a concise description
of the hamiltonian of the model which is needed in this work.
The building blocks of the \mbox{$sdg$-IBM}
are $s$, $d$ and $g$ bosons with angular momenta $\ell=0$, 2 and 4.
A nucleus is characterized by a constant total number of bosons $N$
which equals half the number of valence nucleons
(particles or holes, whichever is smaller).
No distinction is made here between neutron and proton bosons.

Since the hamiltonian of the \mbox{$sdg$-IBM} conserves the total number of bosons,
it can be written in terms of the 225 operators $b_{\ell m}^\dag b_{\ell' m'}$
where $b_{\ell m}^\dag$ ($b_{\ell m}$) creates (annihilates)
a boson with angular momentum $\ell$ and $z$ projection $m$.
This set of 225 operators generates the Lie algebra U(15)
of unitary transformations in fifteen dimensions.
A hamiltonian that conserves the total number of bosons
is of the generic form
\begin{equation}
\hat H=E_0+\hat H^{(1)}+\hat H^{(2)}+\cdots,
\label{e_ham}
\end{equation}
where the index refers to the order of the interaction
in the generators of U(15).
The first term $E_0$ is a constant
which represents the binding energy of the core. 
The second term is the one-body part
\begin{eqnarray}
\hat H^{(1)}&=&
\epsilon_s[s^\dag\times\tilde s]^{(0)}+
\epsilon_d\sqrt{5}[d^\dag\times\tilde d]^{(0)}+
\epsilon_g\sqrt{9}[g^\dag\times\tilde g]^{(0)}
\nonumber\\&\equiv&
\epsilon_s\hat n_s+
\epsilon_s\hat n_d+
\epsilon_d\hat n_g,
\label{e_ham1}
\end{eqnarray}
where $\times$ refers to coupling in angular momentum
(shown as an upperscript in round brackets),
$\tilde b_{\ell m}\equiv(-)^{\ell-m}b_{\ell,-m}$
and the coefficients $\epsilon_s$, $\epsilon_d$ and $\epsilon_g$
are the energies of the $s$, $d$ and $g$ bosons, respectively.
The third term in the hamiltonian~(\ref{e_ham})
represents the two-body interaction
\begin{equation}
\hat H^{(2)}=
\sum_{\ell_1\leq\ell_2,\ell'_1\leq\ell'_2,L}
\tilde v^L_{\ell_1\ell_2\ell'_1\ell'_2}
[[b^\dag_{\ell_1}\times b^\dag_{\ell_2}]^{(L)}\times
[\tilde b_{\ell'_2}\times\tilde b_{\ell'_1}]^{(L)}]^{(0)}_0,
\label{e_ham2}
\end{equation}
where the coefficients $\tilde v$ are related to the interaction matrix elements
between normalized two-boson states, 
\begin{equation}
\tilde v^L_{\ell_1\ell_2\ell'_1\ell'_2}=
\sqrt{\frac{2L+1}{(1+\delta_{\ell_1\ell_2})(1+\delta_{\ell'_1\ell'_2})}}
\langle\ell_1\ell_2;LM|\hat H^{(2)}|\ell'_1\ell'_2;LM\rangle.
\label{e_int}
\end{equation}
Since the bosons are necessarily symmetrically coupled,
allowed two-boson states are
$s^2$ ($L=0$), $sd$ ($L=2$), $sg$ ($L=4$),
$d^2$ ($L=0,2,4$), $dg$ ($L=2,3,4,5,6$)
and $g^2$ ($L=0,2,4,6,8$).
Since for $n$ states with a given angular momentum
one has $n(n+1)/2$ interactions,
32 independent two-body interactions $v$ are found:
six for $L=0$,
ten for $L=2$,
one for $L=3$,
ten for $L=4$,
one for $L=5$,
three for $L=6$
and one for $L=8$.
Together with the constant $E_0$
and the three single-boson energies
$\epsilon_s$, $\epsilon_d$ and $\epsilon_g$,
a total of 36 parameters is thus needed to specify completely
the hamiltonian of the \mbox{$sdg$-IBM}
which includes up to two-body interactions.
This number is far too great for practical applications.
In the following sections we explain how possible simplifications
draw inspiration from the classical limit
and from the \mbox{$sd$-IBM}.

\section{The classical limit}
\label{s_clas}
The classical limit of an IBM hamiltonian
is defined as its expectation value
in a coherent state~\cite{Gilmore79}
and yields a function of the relevant deformation parameters
which can be interpreted as a potential surface
depending on these parameters.
The method was first proposed
for the \mbox{$sd$-IBM}~\cite{Ginocchio80,Dieperink80,Bohr80}.
The extension to the \mbox{$sdg$-IBM}
was carried out by Devi and Kota~\cite{Devi90}
who established the classical limit
of the various limits of U(15).

The coherent state for the \mbox{$sdg$-IBM}
is given by
\begin{equation}
|N;\alpha_{2\mu},\alpha_{4\mu}\rangle\propto
\left(s^\dag+
\sum_\mu\alpha_{2\mu}d^\dag_\mu+
\sum_\mu\alpha_{4\mu}g^\dag_\mu\right)^N|{\rm o}\rangle,
\label{e_coh}
\end{equation}
where $|{\rm o}\rangle$ is the boson vacuum
and the $\alpha_{\lambda\mu}$
are the spherical tensors of quadrupole ($\lambda=2$) 
and hexadecapole ($\lambda=4$) deformation.
They have the interpretation of shape variables
and their associated conjugate momenta.
If one limits oneself to static problems,
the $\alpha_{\lambda\mu}$ can be taken as real;
they specify a shape
and are analogous to the shape variables
of the droplet model of the nucleus~\cite{BM75}.
In particular, they satisfy the same constraints as the latter
which follow from reflection symmetry, that is,
(i) $\alpha_{\lambda\mu}=\alpha_{\lambda-\mu}$
and (ii) $\alpha_{\lambda\mu}=0$ for $\mu$ odd.

The most general parametrization of a surface
with quadrupole and hexadecapole deformation
has been discussed by Rohozi\'nski and Sobiczewski~\cite{Rohozinski81}
and involves five shape parameters:
two of quadrupole ($\beta_2$ and $\gamma_2$)
and three of hexadecapole character ($\beta_4$, $\gamma_4$ and $\delta_4$).
A considerable simplification of the problem is obtained
by relating the two hexadecapole variables ($\gamma_4$ and $\delta_4$)
that parametrize deviations from axial symmetry
in terms of the corresponding quadrupole variable $\gamma_2$,
according to the Cayley-Hamilton theorem~\cite{Nazarewicz81}.
In terms of the latter variables,
the coherent state~(\ref{e_coh}) is rewritten as
\begin{eqnarray}
|N;\beta_2,\beta_4,\gamma_2\rangle&=&
\sqrt{\frac{1}{N!(1+\beta_2^2+\beta_4^2)^N}}
\nonumber\\&\times&
\Bigl(s^\dag+
\beta_2\Bigl[\cos\gamma_2d^\dag_0+
\sqrt{\textstyle{\frac 1 2}}\sin\gamma_2(d^\dag_{-2}+d^\dag_{+2})\Bigr]
\nonumber\\&&+
\textstyle{{\frac 1 6}}\beta_4\Bigl[(5\cos^2\gamma_2+1)g^\dag_0+
\sqrt{\textstyle{\frac{15}{2}}}\sin2\gamma_2(g^\dag_{-2}+g^\dag_{+2})
\nonumber\\&&+
\sqrt{\textstyle{\frac{35}{2}}}\sin^2\gamma_2(g^\dag_{-4}+g^\dag_{+4})\Bigr]
\Bigr)^N|{\rm o}\rangle.
\label{e_cohb}
\end{eqnarray}
The range of allowed values for the quadrupole shape variables is
$0\leq\beta_2<+\infty$ and $0\leq\gamma_2\leq\pi/3$.
On the other hand, the hexadecapole shape variable $\beta_4$
is unrestricted~\cite{Nazarewicz81}.
The expectation value of the hamiltonian~(\ref{e_ham}) in this state
can be determined by elementary methods~\cite{Isacker81}
and yields a function of $\beta_2$, $\beta_4$ and $\gamma_2$
which is identified with a potential $V(\beta_2,\beta_4,\gamma_2)$.
In this way the classical limit of the hamiltonian~(\ref{e_ham})
has the following generic form:
\begin{eqnarray}
V(\beta_2,\beta_4,\gamma_2)&=&
E_0+\sum_{n\geq1}\frac{N(N-1)\cdots(N-n+1)}{(1+\beta_2^2+\beta_4^2)^n}
\nonumber\\&&\qquad\times
\sum_{klm}a^{(n)}_{kl;m}(\beta_2)^k(\beta_4)^l\cos(3m\gamma_2),
\label{e_climit}
\end{eqnarray}
where $n$ refers to the order of the interaction.
If up to two-body terms are included in the hamiltonian,
the non-zero coefficients $a^{(n)}_{kl;m}$ are
\begin{eqnarray}&&
a^{(1)}_{00;0}=\epsilon_s,
\quad
a^{(1)}_{20;0}=\epsilon_d,
\quad
a^{(1)}_{02;0}=\epsilon_g,
\nonumber\\&&
\textstyle
a^{(2)}_{00;0}={\frac 1 2}v_{ssss}^0,
\quad
a^{(2)}_{20;0}=\sqrt{\frac 1 5}v_{ssdd}^0+v_{sdsd}^2,
\quad
a^{(2)}_{02;0}={\frac 1 3}v_{ssgg}^0+v_{sgsg}^4,
\nonumber\\&&
\textstyle
a^{(2)}_{30;1}=-2\sqrt{\frac 1 7}v_{sddd}^2,
\quad
a^{(2)}_{21;0}=2\sqrt{\frac 2 7}v_{sddg}^2+\frac{6}{\sqrt{35}}v_{sgdd}^4,
\nonumber\\&&
\textstyle
a^{(2)}_{12;1}=-\frac{10}{3}\sqrt{\frac{2}{77}}v_{sdgg}^2-4\sqrt{\frac{5}{77}}v_{sgdg}^4,
\quad
a^{(2)}_{03;0}=-\frac{112}{9\sqrt{1001}}v_{sggg}^4,
\nonumber\\&&
\textstyle
a^{(2)}_{03;2}=-\frac{50}{9\sqrt{1001}}v_{sggg}^4,
\quad
a^{(2)}_{40;0}=
\frac{1}{10}v_{dddd}^0+{\frac 1 7}v_{dddd}^2+\frac{9}{35}v_{dddd}^4,
\nonumber\\&&
\textstyle
a^{(2)}_{31;1}=
-\frac{2\sqrt{2}}{7}v_{dddg}^2-
\frac{12}{7\sqrt{11}}v_{dddg}^4,
\nonumber\\&&
\textstyle
a^{(2)}_{22;0}=
\frac{1}{3\sqrt{5}}v_{ddgg}^0+
\frac{5}{9}\sqrt{\frac{2}{11}}v_{ddgg}^2+
\frac{16}{3\sqrt{715}}v_{ddgg}^4
\nonumber\\&&
\textstyle
\phantom{a^{(2)}_{22;0}=}+
\frac{2}{7}v_{dgdg}^2+
\frac{1}{18}v_{dgdg}^3+
\frac{85}{462}v_{dgdg}^4+
\frac{47}{99}v_{dgdg}^6,
\nonumber\\&&
\textstyle
a^{(2)}_{22;2}=
\frac{5}{63}\sqrt{\frac{2}{11}}v_{ddgg}^2+
\frac{10}{21}\sqrt{\frac{5}{143}}v_{ddgg}^4-
\frac{1}{18}v_{dgdg}^3+
\frac{5}{66}v_{dgdg}^4-
\frac{2}{99}v_{dgdg}^6,
\nonumber\\&&
\textstyle
a^{(2)}_{13;1}=-
\frac{20}{21\sqrt{11}}v_{dggg}^2-
\frac{36}{77}\sqrt{\frac{5}{13}}v_{dggg}^4-
\frac{10\sqrt{2}}{33}v_{dggg}^6,
\nonumber\\&&
\textstyle
a^{(2)}_{04;0}=
\frac{1}{18}v_{gggg}^0+
\frac{625}{6237}v_{gggg}^2+
\frac{167}{2457}v_{gggg}^4+
\frac{5}{81}v_{gggg}^6+
\frac{2485}{11583}v_{gggg}^8,
\nonumber\\&&
\textstyle
a^{(2)}_{04;2}=
\frac{25}{891}v_{gggg}^2+
\frac{50}{3861}v_{gggg}^4+
\frac{35}{891}v_{gggg}^6+
\frac{280}{11583}v_{gggg}^8,
\label{e_coeff}
\end{eqnarray}
in terms of the single-boson energies $\epsilon_\ell$
and the matrix elements between normalized two-boson states,
\begin{equation}
v^L_{\ell_1\ell_2\ell'_1\ell'_2}\equiv
\langle\ell_1\ell_2;LM|\hat H^{(2)}|\ell'_1\ell'_2;LM\rangle.
\label{e_intb}
\end{equation}

This analysis shows that the geometry of the $sdg$ hamiltonian
depends on 19 parameters (including $E_0$),
considerably smaller than the 36 parameters
of the quantum-mechanical version of the same hamiltonian.
The structure of the phase diagram associated
with the general $sdg$ hamiltonian~(\ref{e_climit})
might be of interest but is well beyond the scope of the present analysis.
In the next section we show
how a simplified $sdg$ hamiltonian can be constructed
which presumably captures the essential features
of the phase-transitional behaviour in the \mbox{$sdg$-IBM}.

\section{A generalized Casten triangle}
\label{s_cast}
A simplified hamiltonian of the \mbox{$sd$-IBM}
is of the form
\begin{equation}
\hat H_{sd}=
\epsilon_d\,\hat n_d+
\kappa\,\hat Q\cdot\hat Q+
\kappa'\hat L\cdot\hat L,
\label{e_cassd}
\end{equation}
where $\hat Q$ is the quadrupole operator with components
$\hat Q_\mu\equiv
[d^\dag\times\tilde s+s^\dag\times\tilde d]^{(2)}_\mu+
\chi[d^\dag\times\tilde d]^{(2)}_\mu$
and $\hat L$ is the angular momentum operator,
$\hat L_\mu\equiv\sqrt{10}\,[d^\dag\times\tilde d]^{(1)}_\mu$.
The $\hat Q^2$ and $\hat L^2$ terms in~(\ref{e_cassd})
constitute the hamiltonian
of the so-called consistent-$Q$ formalism (CQF)~\cite{Warner82};
for $\chi=\pm\sqrt{7}/2$ it gives rise to the deformed or SU(3) limit 
and for $\chi=0$ to the $\gamma$-unstable or SO(6) limit.
In an extended consistent-$Q$ formalism (ECQF)~\cite{Lipas85}
the term $\hat n_d$ is added
with which the third, vibrational or U(5) limit of the \mbox{$sd$-IBM}
can be obtained.
The ECQF hamiltonian thus allows one
to reach all three limits of the model with four parameters.

The eigenfunctions of the ECQF hamiltonian
are, in fact, independent of $\kappa'$
and, furthermore, of the overall scale of the spectrum,
reducing the number of essential parameters to two,
the ratio $\kappa/\epsilon_d$ and $\chi$.
The physically relevant portion of this parameter space
can be represented on a so-called Casten triangle~\cite{Casten81}.
Its vertices correspond to the three limits of the \mbox{$sd$-IBM}, U(5), SU(3) and SO(6)
for spherical, deformed and $\gamma$-unstable nuclei, respectively.
The three legs of the triangle describe transitional cases,
intermediate between two of the three limits,
while an arbitrary point on the triangle
corresponds to an admixture of the three limits.
 
In this paper we propose a similar hamiltonian in the \mbox{$sdg$-IBM}.
The different dynamical symmetries of the \mbox{$sdg$-IBM}
are known since long~\cite{Demeyer86,Kota87}.
In particular, three limits occur in \mbox{$sdg$-IBM},
namely ${\rm U}(5)\otimes{\rm U}(9)$, SU(3) and SO(15),
that are the analogues of the limits of the \mbox{$sd$-IBM}.
An immediate generalization of the $sd$ hamiltonian~(\ref{e_cassd})
is therefore of the form
\begin{equation}
\hat H_{sdg}=
\epsilon_d\hat n_d+
\epsilon_g\hat n_g-
\kappa\hat Q^{(2)}\cdot\hat Q^{(2)}-
\kappa(1-\chi^2)\hat Q^{(4)}\cdot\hat Q^{(4)},
\label{e_cassdg}
\end{equation}
where $\hat Q^{(2)}_\mu$ and $\hat Q^{(4)}_\mu$
are the quadrupole and hexadecapole operators, respectively,
given by
\begin{eqnarray}
\hat Q^{(2)}_\mu&=&
\sigma[s^\dag\times\tilde d+d^\dag\times\tilde s]^{(2)}_\mu
\nonumber\\&&+
\textstyle
\chi\left(
\frac{11\sqrt{10}}{28}[d^\dag\times\tilde d]^{(2)}_\mu-
\frac{9}{7}\sigma'[d^\dag\times\tilde g+g^\dag\times\tilde d]^{(2)}_\mu+
\frac{3\sqrt{55}}{14}[g^\dag\times\tilde g]^{(2)}_\mu\right),              
\nonumber\\
\hat Q^{(4)}_\mu&=&[s^\dag\times\tilde g+g^\dag\times\tilde s]^{(4)}_\mu.
\label{e_q24}
\end{eqnarray}
The parameter $\chi$ is chosen so that $0\leq\chi\leq1$
and $\sigma,\sigma'=\pm1$ are two arbitrary phases
leading to four possible realizations of SU(3)~\cite{Kota87}.
For $\chi=1$ the quadrupole operator
is a generator of \mbox{$sdg$-SU(3)}
while for $\chi=0$ it becomes a generator of SO(15).
Other parametrizations of the hexadecapole strength with the same effect
could be taken in the hamiltonian~(\ref{e_cassdg}),
such as $\sqrt{1-\chi^2}$ or $(1-|\chi|)$;
they differ little in actual results
and the dependence $(1-\chi^2)$ is chosen here for its convenience. 
Note that with this parametrization
the deformed limit has a vanishing hexadecapole interaction
which is then only present in the $\gamma$-soft SO(15) limit.
To do otherwise would require a hexadecapole operator
taken from one of the other two strong coupling limits of \mbox{$sdg$-IBM},
SU(5) or SU(6).
Since these have a questionable microscopic interpretation~\cite{Bouldjedri05},
this is not done here.

With the help of the expression~(\ref{e_coeff})
for the general hamiltonian of the \mbox{$sdg$-IBM}
the classical limit of the special hamiltonian~(\ref{e_cassdg}) yields
\begin{eqnarray}
\frac{\langle\hat H_{sdg}\rangle}{N}&=&
\frac{\epsilon_d\beta_2^2}{1+\beta_2^2+\beta_4^2}+
\frac{\epsilon_g\beta_4^2}{1+\beta_2^2+\beta_4^2}-
\frac{(N-1)\kappa}{(1+\beta_2^2+\beta_4^2)^2}
\nonumber\\&&\times
\textstyle
\Biggl[4\beta_2^2-
\sigma\sigma'\chi\Bigl(
\frac{72}{7}\sqrt{\frac 2 7}\beta_2^2\beta_4+
\frac{2}{7}\sqrt{\frac 5 7}\bigl[11\beta_2^3+10\beta_2\beta_4^2\bigr]
\sigma'\cos3\gamma_2\Bigr)
\nonumber\\&&\phantom{\times}+
\textstyle
\chi^2\Bigl(
\frac{605}{1372}\beta_2^4+
\frac{5813}{2058}\beta_2^2\beta_4^2+
\frac{3125}{6174}\beta_4^4+
\frac{18\sqrt{10}}{343}\bigl[11\beta_2^3\beta_4
\nonumber\\&&\phantom{\times\chi^2\Bigl(}+
\textstyle
10\beta_2\beta_4^3\bigr]
\sigma'\cos3\gamma_2-
\frac{25}{6174}\bigl[33\beta_2^2\beta_4^2+35\beta_4^4\bigr]
\cos6\gamma_2\Bigr)\Biggr]
\nonumber\\&&-
\frac{4(N-1)\kappa(1-\chi^2)\beta_4^2}{(1+\beta_2^2+\beta_4^2)^2},
\label{e_clim}
\end{eqnarray}
where only the two-body parts
of the $\hat Q^{(2)}\cdot\hat Q^{(2)}$
and $\hat Q^{(4)}\cdot\hat Q^{(4)}$ operators
have been considered.
We note first of all that a change of sign of the product $\sigma\sigma'$
is equivalent with the change $\chi\rightarrow-\chi$.
We may thus take $\sigma\sigma'=+1$
and consider the parameter range $-1\leq\chi\leq+1$.
Secondly, we remark that everywhere else $\sigma'$
occurs in combination with $\cos3\gamma_2$
and, consequently, the change $\sigma'\rightarrow-\sigma'$
is equivalent with the replacement $\gamma_2\rightarrow\pi/3-\gamma_2$.
This means that without loss of generality
the choice $\sigma=\sigma'=+1$ can be made
as long as the range $-1\leq\chi\leq+1$ is considered.
This can also be understood from the form
of the $\hat Q^{(2)}_\mu$ operator in eq.~(\ref{e_q24})
from which it is clear that the change $\sigma\rightarrow-\sigma$
is equivalent to $\chi\rightarrow-\chi$.
The two choices $\chi=-1$ and $\chi=+1$ each
correspond to a dynamical-symmetry limit of the SU(3) type
which shall be referred to as ${\rm SU}_-(3)$ and ${\rm SU}_+(3)$, respectively.

Choosing $\epsilon_d$ as the overall energy scale,
we summarize the preceding discussion
by stating that we seek to minimize (in $\beta_2$, $\beta_4$ and $\gamma_2$)
the following energy surface:
\begin{eqnarray}
\lefteqn{V_3(\beta_2,\beta_4,\gamma_2;\eta,\chi,r)}
\nonumber\\&=&
\frac{\beta_2^2}{1+\beta_2^2+\beta_4^2}+
\frac{r\beta_4^2}{1+\beta_2^2+\beta_4^2}-
\frac{\eta}{(1+\beta_2^2+\beta_4^2)^2}
\nonumber\\&&\times
\textstyle
\Biggl[4\beta_2^2-
\chi\Bigl(
\frac{72}{7}\sqrt{\frac 2 7}\beta_2^2\beta_4+
\frac{2}{7}\sqrt{\frac 5 7}\bigl[11\beta_2^3+10\beta_2\beta_4^2\bigr]
\cos3\gamma_2\Bigr)
\nonumber\\&&\phantom{\times}+
\textstyle
\chi^2\Bigl(
\frac{605}{1372}\beta_2^4+
\frac{5813}{2058}\beta_2^2\beta_4^2+
\frac{3125}{6174}\beta_4^4+
\frac{18\sqrt{10}}{343}\bigl[11\beta_2^3\beta_4
\nonumber\\&&\phantom{\times\chi^2\Bigl(}+
\textstyle
10\beta_2\beta_4^3\bigr]
\cos3\gamma_2-
\frac{25}{6174}\bigl[33\beta_2^2\beta_4^2+35\beta_4^4\bigr]
\cos6\gamma_2\Bigr)\Biggr]
\nonumber\\&&-
\frac{4\eta(1-\chi^2)\beta_4^2}{(1+\beta_2^2+\beta_4^2)^2},
\label{e_v3}
\end{eqnarray}
where the notation $\eta\equiv(N-1)\kappa/\epsilon_d$
and $r\equiv\epsilon_g/\epsilon_d$ is introduced.
The order parameters are $\beta_2$, $\beta_4$ and $\gamma_2$,
and the control parameters are $\eta$, $\chi$ and $r$.
The subscript 3 in $V_3$ refers to the number of order parameters
and is used to distinguish it from $V_2$ introduced below.
Physically meaningful values of the parameters imply $\eta>0$.
It is convenient to introduce, instead of $\eta=\xi/4(1-\xi)$,
the control parameter $\xi=4\eta/(1+4\eta)$
such that $\xi\in[0,1]$ instead of $\eta\in[0,+\infty[$.
(The combination $4\eta$ instead of $\eta$ is chosen in $\xi$ for later convenience.)
Furthermore, since the $g$ boson
is higher in energy than the $d$ boson we have $r>1$.
In the limit $r\rightarrow\infty$ one recovers the \mbox{$sd$-IBM}.
Again, it is more convenient to introduce, instead of $r=1/(1-\zeta)$,
the control parameter $\zeta=(r-1)/r$
such that $\zeta\in[0,1]$ instead of $r\in[1,\infty[$.
The definition adopted here for $\eta$ or $\xi$
is similar to the one of Iachello~\cite{Iachello87},
so that we can easily compare with \mbox{$sd$-IBM} for $\zeta=1$.

\begin{figure}
\centering
\includegraphics[width=8cm]{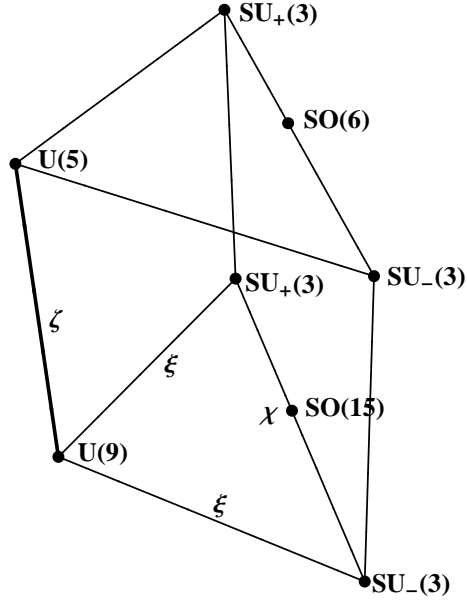}
\caption{The Casten prism associated
with the hamiltonian~(\ref{e_cassdg}) of the \mbox{$sdg$-IBM}.
The different symmetry limits
of the \mbox{$sd$-IBM} and \mbox{$sdg$-IBM}
are indicated with black dots.
The three control parameters $\xi$, $\chi$ and $\zeta$, as defined in the text,
are indicated alongside the corresponding axes of the prism.}
\label{f_prism}
\end{figure}

The parameter space covered by the hamiltonian~(\ref{e_cassdg})
can be summarized with a generalization of the Casten triangle
which becomes a prism, as shown in figure~\ref{f_prism}.
Also indicated on the figure are the different symmetry limits
of the \mbox{$sd$-IBM} and \mbox{$sdg$-IBM}.
The point $(\xi,\zeta)=(0,1)$ is denoted as U(5), as should be,
and the point $(\xi,\zeta)=(0,0)$, somewhat arbitrarily, as U(9),
so that all vertices of the prism are labelled,
facilitating the discussion of the structure of the phase diagram.
Note that the ${\rm U}(5)\otimes{\rm U}(9)$ symmetry
is valid for any combination of $\epsilon_d$ and $\epsilon_g$
(and hence for all $\zeta\in[0,1]$) as long as $\xi=0$.

\section{Method}
\label{s_meth}
The problem here is more complicated than in \mbox{$sd$-IBM}
because of an extra order parameter ($\beta_4$)
and an extra control parameter ($r$).
The necessary and sufficient conditions for the potential energy $V$
to have a (Morse or normal) critical point in $(\beta_2^*,\beta_4^*,\gamma_2^*)$ are
\begin{equation}
\left.\frac{\partial V_3}{\partial\beta_2}\right|_{(\beta_2^*,\beta_4^*,\gamma_2^*)}=
\left.\frac{\partial V_3}{\partial\beta_4}\right|_{(\beta_2^*,\beta_4^*,\gamma_2^*)}=
\left.\frac{\partial V_3}{\partial\gamma_2}\right|_{(\beta_2^*,\beta_4^*,\gamma_2^*)}=0.
\label{e_crit3}
\end{equation}
Furthermore, whether the critical point is a minimum, maximum or saddle point
depends on whether the eigenvalues of the stability matrix
in $(\beta_2^*,\beta_4^*,\gamma_2^*)$,
\begin{equation}
{\cal H}_3(\beta_2^*,\beta_4^*,\gamma_2^*;\xi,\chi,\zeta)\equiv
\left[\begin{array}{ccc}
\displaystyle\frac{\partial^2V_3}{\partial\beta_2^2}&
\displaystyle\frac{\partial^2V_3}{\partial\beta_2\partial\beta_4}&
\displaystyle\frac{\partial^2V_3}{\partial\beta_2\partial\gamma_2}\\[2ex]
\displaystyle\frac{\partial^2V_3}{\partial\beta_4\partial\beta_2}&
\displaystyle\frac{\partial^2V_3}{\partial\beta_4^2}&
\displaystyle\frac{\partial^2V_3}{\partial\beta_4\partial\gamma_2}\\[2ex]
\displaystyle\frac{\partial^2V_3}{\partial\gamma_2\partial\beta_2}&
\displaystyle\frac{\partial^2V_3}{\partial\gamma_2\partial\beta_4}&
\displaystyle\frac{\partial^2V_3}{\partial\gamma_2^2}
\end{array}\right]_{(\beta_2^*,\beta_4^*,\gamma_2^*)},
\label{e_stab3}
\end{equation}
are all positive, all negative or both positive and negative.
If, in addition to the condition~(\ref{e_crit3}),
the determinant of the stability matrix vanishes,
\begin{equation}
\Delta_3(\beta_2^*,\beta_4^*,\gamma_2^*;\xi,\chi,\zeta)\equiv
\det{\cal H}_3(\beta_2^*,\beta_4^*,\gamma_2^*;\xi,\chi,\zeta)=0,
\label{e_deg3}
\end{equation}
a so-called degenerate (or non-Morse) critical point is found~\cite{Gilmore81}.
This ensemble is of particular importance
since it defines the points
at which the energy surface changes in structure
and hence at which the system possibly undergoes a shape phase transition.
The points that satisfy the condition~(\ref{e_deg3})
define a surface in the control parameters space $(\xi,\chi,\zeta)$.
Degenerate critical points shall be denoted as $(\xi^*,\chi^*,\zeta^*)$.
Our goal will be to determine the appearance of the degenerate critical surface
in the Casten prism of figure~\ref{f_prism}.

The problem of determining the phase diagram for the Casten prism
can be addressed by noting that solutions of the last of the equations in~(\ref{e_crit3})
satisfy one of the two equations
\begin{eqnarray}
\sin3\gamma_2=0,
\qquad
\cos3\gamma_2=
\frac{9\beta_2(9\sqrt{2}\chi\beta_4-7\sqrt{7})(11\beta_2^2+10\beta_4^2)}
{5\sqrt{5}\chi\beta_4^2(33\beta_2^2+35\beta_4^2)}.
\label{e_solg2}
\end{eqnarray}
These two conditions define the two possible classes of solutions.
The {\em first class} corresponds to critical points with axial symmetry,
$\gamma_2^*=0$ (prolate) or $\gamma_2^*=\pi/3$ (oblate).
A further simplification is then obtained by noting that
the off-diagonal elements of the stability matrix involving $\gamma_2$
vanish at $\gamma_2^*=0$ or $\gamma_2^*=\pi/3$,
and hence the points in the degenerate critical set $(\xi^*,\chi^*,\zeta^*)$
satisfy one of the two following equations:
\begin{equation}
\det\left[\begin{array}{ccc}
\displaystyle\frac{\partial^2V_3}{\partial\beta_2^2}&
\displaystyle\frac{\partial^2V_3}{\partial\beta_2\partial\beta_4}\\[2ex]
\displaystyle\frac{\partial^2V_3}{\partial\beta_4\partial\beta_2}&
\displaystyle\frac{\partial^2V_3}{\partial\beta_4^2}\\
\end{array}\right]_{(\beta_2^*,\beta_4^*,\gamma_2^*)}=0,
\qquad
\left.\frac{\partial^2V_3}{\partial\gamma_2^2}\right|_{(\beta_2^*,\beta_4^*,\gamma_2^*)}=0.
\end{equation}
Furthermore, we note that $\cos3\gamma_2$ always occurs in Eq.~(\ref{e_v3})
in combination with an odd power (1 or 3) of $\beta_2$.
Since the replacement $\gamma_2\rightarrow\pi/3-\gamma_2$
corresponds to a change in sign of $\cos3\gamma_2$
and no change in $\cos6\gamma_2$,
both cases $\gamma_2=0$ and $\gamma_2=\pi/3$ are covered
by putting $\gamma_2=0$
and allowing also negative values of $\beta_2$.
This further simplifies the potential energy surface~(\ref{e_v3})
which now only depends on the two order parameters $\beta_2$ and $\beta_4$,
\begin{eqnarray}
\lefteqn{V_2(\beta_2,\beta_4;\eta,\chi,r)}
\nonumber\\&=&
\frac{\beta_2^2}{1+\beta_2^2+\beta_4^2}+
\frac{r\beta_4^2}{1+\beta_2^2+\beta_4^2}-
\frac{\eta}{(1+\beta_2^2+\beta_4^2)^2}
\nonumber\\&&\times
\textstyle
\Biggl[4\beta_2^2+
\chi\Bigl(
\frac{22}{7}\sqrt{\frac 5 7}\beta_2^3+
\frac{72}{7}\sqrt{\frac 2 7}\beta_2^2\beta_4+
\frac{20}{7}\sqrt{\frac 5 7}\beta_2\beta_4^2\Bigr)
\nonumber\\&&\phantom{\times}+
\textstyle
\chi^2\Bigl(
\frac{605}{1372}\beta_2^4+
\frac{198\sqrt{10}}{343}\beta_2^3\beta_4+
\frac{923}{343}\beta_2^2\beta_4^2+
\frac{180\sqrt{10}}{343}\beta_2\beta_4^3+
\frac{125}{343}\beta_4^4\Bigr)\Biggr]
\nonumber\\&&-
\frac{4\eta(1-\chi^2)\beta_4^2}{(1+\beta_2^2+\beta_4^2)^2}.
\label{e_v2}
\end{eqnarray}
In summary, critical points $(\beta_2^*,\beta_4^*)$ with axial symmetry
are obtained from the simultaneous conditions
\begin{equation}
\left.\frac{\partial V_2}{\partial\beta_2}\right|_{(\beta_2^*,\beta_4^*)}=
\left.\frac{\partial V_2}{\partial\beta_4}\right|_{(\beta_2^*,\beta_4^*)}=0,
\label{e_crit2}
\end{equation}
where one assumes $\gamma_2^*=0$ and allows negative values of $\beta_2^*$.
The points $(\xi^*,\chi^*,\zeta^*)$ in the degenerate critical set
should satisfy one of the two following equations:
\begin{equation}
\Delta_2(\beta_2^*,\beta_4^*;\xi,\chi,\zeta)=0,
\qquad
\left.\frac{\partial^2V_3}{\partial\gamma_2^2}\right|_{(\beta_2^*,\beta_4^*,\gamma_2^*)}=0,
\label{e_deg2}
\end{equation}
where $\Delta_2(\beta_2^*,\beta_4^*;\xi,\chi,\zeta)$
is the determinant of the $2\times2$ stability matrix,
\begin{equation}
{\cal H}_2(\beta_2^*,\beta_4^*;\xi,\chi,\zeta)\equiv
\left[\begin{array}{ccc}
\displaystyle\frac{\partial^2V_2}{\partial\beta_2^2}&
\displaystyle\frac{\partial^2V_2}{\partial\beta_2\partial\beta_4}\\[2ex]
\displaystyle\frac{\partial^2V_2}{\partial\beta_4\partial\beta_2}&
\displaystyle\frac{\partial^2V_2}{\partial\beta_4^2}\\
\end{array}\right]_{(\beta_2^*,\beta_4^*)}.
\label{e_stab2}
\end{equation}

The {\em second class} corresponds to critical points without axial symmetry,
obtained as solutions of the second equation in~(\ref{e_solg2}).

Before attacking the problem of the occurrence of shape phase transitions
for the entire hamiltonian~(\ref{e_cassdg}),
it is instructive to study, as a function of $r$ or $\zeta$, the three legs of the triangle
between the limits ${\rm U}(5)\otimes{\rm U}(9)$, ${\rm SU}_+(3)$ and ${\rm SU}_-(3)$.
This will determine the structure of the phase diagram
on the faces of the prism of figure~\ref{f_prism}. 
In this analysis care should be taken to cover the two cases $\chi>0$ and $\chi<0$,
and study in particular the behaviour at $\chi=0$,
for instance in the ${\rm SU}_-(3)$--SO(15)--${\rm SU}_+(3)$ transition.

\section{Partial phase diagrams}
\label{s_part}

\subsection{The U(5) $\otimes$ U(9) to SO(15) transition}
\label{ss_u59so15}
The ${\rm U}(5)\otimes{\rm U}(9)$--SO(15) transition is obtained
for $\chi=0$ and $r=1$ (or $\zeta=0$)
while $\eta$ varies from 0 to $+\infty$ (or $\xi$ from 0 to 1).
At the top side of the prism, for $r=\infty$ (or $\zeta=1$),
this reduces to the U(5)--SO(6) transition.
We are interested here in the entire phase diagram
in $\eta$ and $r$ (or $\xi$ and $\zeta$)
which can be established by an expansion
in $\beta_2$ and $\beta_4$ around (0,0).
Since there is no dependence on $\gamma_2$ for $\chi=0$,
the analysis can be done starting from the potential~(\ref{e_v2}).
Up to fourth order we obtain
\begin{eqnarray}
V_2(\beta_2,\beta_4;\eta,\chi=0,r)&\approx&
(1-4\eta)\beta_2^2+
(r-4\eta)\beta_4^2-
(1-8\eta)\beta_2^4
\nonumber\\&&-
(1-16\eta+r)\beta_2^2\beta_4^2-
(r-8\eta)\beta_4^4,
\label{e_exp1}
\end{eqnarray}
involving even powers of $\beta_2$ and $\beta_4$ only.
The expansion~(\ref{e_exp1}) shows that
the potential always exhibits a minimum, saddle point or maximum
at $\beta_2^*=\beta_4^*=0$.
This is confirmed by computing the stability matrix
at $\beta_2^*=\beta_4^*=0$:
\begin{equation}
{\cal H}_2(0,0;\xi,\chi=0,\zeta)=
\left[\begin{array}{cc}
\displaystyle
\frac{2(1-2\xi)}{1-\xi}&0\\[2ex]
0&\displaystyle\frac{2[1-\xi(2-\zeta)]}{(1-\xi)(1-\zeta)}
\end{array}\right].
\label{e_stab10}
\end{equation}
The zeros of the determinant of the stability matrix
determine the degenerate critical points
which separate the different possible shapes of the potential.
They are $\xi=1/2$ (or $\eta=1/4$)
and $\xi=1/(2-\zeta)$ (or $\eta=r/4$).
Consequently, we find that the potential has
(I) a minimum for $\xi<1/2$,
(IIa) a saddle point for $1/2<\xi<1/(2-\zeta)$
and (IIb) a maximum for $\xi>1/(2-\zeta)$,
always at $\beta_2^*=\beta_4^*=0$.

In addition, for $\xi>1/2$, critical points in $\beta_2$
occur for $\beta_2^*\neq0$ ({\it i.e.} quadrupole-deformed critical points).
The dependence of $\beta_2^*$ on $\xi$ and $\zeta$ is obtained
by solving $\partial V_2/\partial\beta_2=0$ for $\beta_4=0$, yielding
\begin{equation}
\beta_2^*=\pm\sqrt{2\xi-1}=
\pm\sqrt{\frac{4-\epsilon_d}{4+\epsilon_d}}.
\label{e_b2}
\end{equation}
Hence the picture emerges
that the spherical minimum $\beta_2^*=\beta_4^*=0$
at $\xi=1/2$ continuously evolves
into the two quadrupole-deformed minima~(\ref{e_b2}).
The phase transition is of second order.
It should be emphasized
that this result is {\em independent} of $\zeta$ (or $r$)
and that the entire line $\xi=1/2$
exhibits a second-order phase transition when crossed.
In particular, we recover the results of \mbox{$sd$-IBM}~\cite{Iachello87}
which corresponds to $\zeta=1$.
One should verify the nature of the critical points
by calculating the stability matrix
at $\beta_2^*=\pm\sqrt{2\xi-1}$ and $\beta_4^*=0$:
\begin{equation}
{\cal H}_2(\pm\sqrt{2\xi-1},0;\xi,\chi=0,\zeta)=
\left[\begin{array}{cc}
\displaystyle
\frac{2\xi-1}{2(1-\xi)\xi^3}&0\\[2ex]
0&\displaystyle\frac{\zeta}{\xi(1-\zeta)}
\end{array}\right].
\label{e_stab1b2}
\end{equation}
Both matrix elements are positive definite
and hence one always has a minimum.

Finally, for $\xi>1/(2-\zeta)$, additional critical points with $\beta_4^*\neq0$ occur
({\it i.e.} hexadecapole-deformed critical points).
The dependence of $\beta_4^*$ on $\xi$ and $\zeta$ is obtained
by solving $\partial V_2/\partial\beta_4=0$ for $\beta_2=0$, yielding
\begin{equation}
\beta_4^*=\pm\sqrt{\frac{\xi(2-\zeta)-1}{1-\xi\zeta}}=
\pm\sqrt{\frac{4-\epsilon_g}{4+\epsilon_g}}.
\label{e_b4}
\end{equation}
So the spherical saddle point $\beta_2^*=\beta_4^*=0$
at $\xi=1/(2-\zeta)$ continuously evolves into the two deformed critical points~(\ref{e_b4}),
the nature of which follows from the stability matrix
at $\beta_2^*=0$ and $\beta_4^*=\pm\sqrt{[\xi(2-\zeta)-1]/(1-\xi\zeta)}$:
\begin{eqnarray}
\lefteqn{{\cal H}_2(0,\pm\sqrt{\frac{\xi(2-\zeta)-1}{1-\xi\zeta}};\xi,\chi=0,\zeta)}
\nonumber\\&&=
\left[\begin{array}{cc}
\displaystyle
-\frac{\zeta(1-\xi\zeta)}{\xi(1-\zeta)^2}&0\\[2ex]
0&\displaystyle\frac{[\xi(2-\zeta)-1](1-\xi\zeta)^3}{2(1-\xi)\xi^3(1-\zeta)^4}
\end{array}\right].
\label{e_stab1b4}
\end{eqnarray}
One matrix element is positive definite
while the other is negative definite
and hence both critical points are saddle points in this case.

\begin{figure}
\centering
\includegraphics[width=10cm]{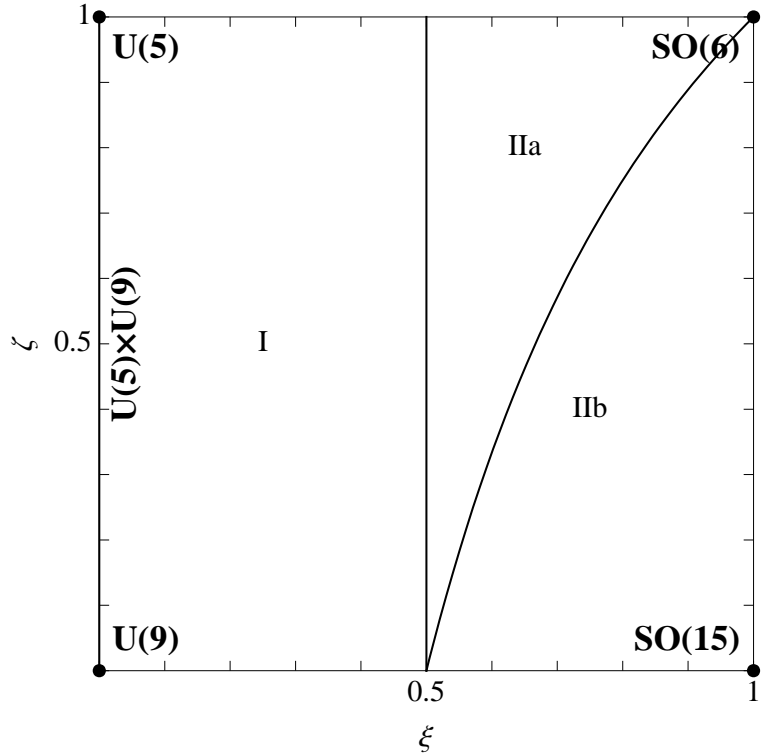}
\caption{The phase diagram in the $(\xi,\zeta)$ plane of the Casten prism for $\chi=0$.
The lines or curves represent the locus of degenerate critical points
that separate the different regions as discussed in the text.
Symmetry limits of the \mbox{$sd$-IBM} and \mbox{$sdg$-IBM}
are indicated with black dots.
The entire $\zeta$ axis
corresponds to the ${\rm U}(5)\otimes{\rm U}(9)$ limit of \mbox{$sdg$-IBM}.}
\label{f_u59so15}
\end{figure}
The phase diagram in the $(\xi,\zeta)$ plane with $\chi=0$
deduced in this way is shown in figure~\ref{f_u59so15}.
The parameter space is divided in three areas
with qualitatively different potentials in $\beta_2$ and $\beta_4$,
characterized by the number and nature (minima, maxima or saddle)
of Morse critical points.
The potential surfaces in the regions
indicated in figure~\ref{f_u59so15}
are characterized by
\begin{itemize}
\item
(I) a spherical minimum ($\beta_2^*=\beta_4^*=0$);
\item
(IIa) a spherical saddle point
and two quadrupole-deformed minima ($\beta_2^*\neq0$, $\beta_4^*=0$);
\item
(IIb) a spherical maximum,
two quadrupole-deformed minima
and two hexadecapole-deformed saddle points ($\beta_2^*=0$, $\beta_4^*\neq0$).
\end{itemize}
The degenerate critical lines in the phase diagram
correspond to {\em bifurcation} lines:
at $\xi=1/2$ the spherical minimum
splits into two quadrupole-deformed minima
and at $\xi=1/(2-\zeta)$ the spherical saddle point
splits into two hexadecapole-deformed saddle points.

\subsection{The U(5) $\otimes$ U(9) to SU(3) transitions}
\label{ss_u59su3}
The ${\rm U}(5)\otimes{\rm U}(9)$--${\rm SU}_+(3)$ transition is obtained
for $\chi=+1$ and $r=1$ (or $\zeta=0$)
while $\eta$ varies from 0 to $+\infty$ (or $\xi$ from 0 to 1).
A similar transition, from ${\rm U}(5)\otimes{\rm U}(9)$ to ${\rm SU}_-(3)$,
is obtained for $\chi=-1$.
At the top side of the prism, for $r=\infty$ (or $\zeta=1$),
the ${\rm U}(5)\otimes{\rm U}(9)$--${\rm SU}_\pm(3)$ transitions
almost reduce to U(5)--${\rm SU}_\pm(3)$,
their analogues in \mbox{$sd$-IBM}.
This correspondence is not exact
since the coefficient of the $[d^\dag\times\tilde d]^{(2)}_\mu$ term
in the quadrupole operator~(\ref{e_q24}) is $11\sqrt{10}/28\approx1.24$
while it is $\sqrt{7}/2\approx1.32$ in \mbox{$sd$-IBM}.
From the expression of the potential surface~(\ref{e_v3})
we note that the change $\chi\rightarrow-\chi$
is equivalent to the simultaneous change $\beta_4\rightarrow-\beta_4$
and $\gamma_2\rightarrow\pi/3-\gamma_2$.
Hence all potential energy surfaces that can be realized for $\chi<0$
follow from those with $\chi>0$
and only one case needs to be studied.
We are interested here in the entire phase diagram
in $\eta$ and $r$ (or $\xi$ and $\zeta$)
and consider the case $\chi=-1$.

An expansion up to third order around $(\beta_2,\beta_4)=(0,0)$ gives
\begin{eqnarray}
V(\beta_2,\beta_4,\gamma_2;\eta,\chi=-1,r)
&\approx&
\textstyle
(1-4\eta)\beta_2^2
+r\beta_4^2
-\frac{22}{7}\sqrt{\frac 5 7}\eta\beta_2^3\cos3\gamma_2
\nonumber\\&&\textstyle
-\frac{72}{7}\sqrt{\frac 2 7}\eta\beta_2^2\beta_4
-\frac{20}{7}\sqrt{\frac 5 7}\eta\beta_2\beta_4^2\cos3\gamma_2.
\label{e_u59su3_exp}
\end{eqnarray}
This shows that the spherical point $(\beta_2,\beta_4)=(0,0)$
is always a minimum in $\beta_4$ since $r>0$
and that it is a minimum in $\beta_2$ for $\xi<1/2$ (or $\eta<1/4$).

\begin{figure}
\centering
\includegraphics[width=10cm]{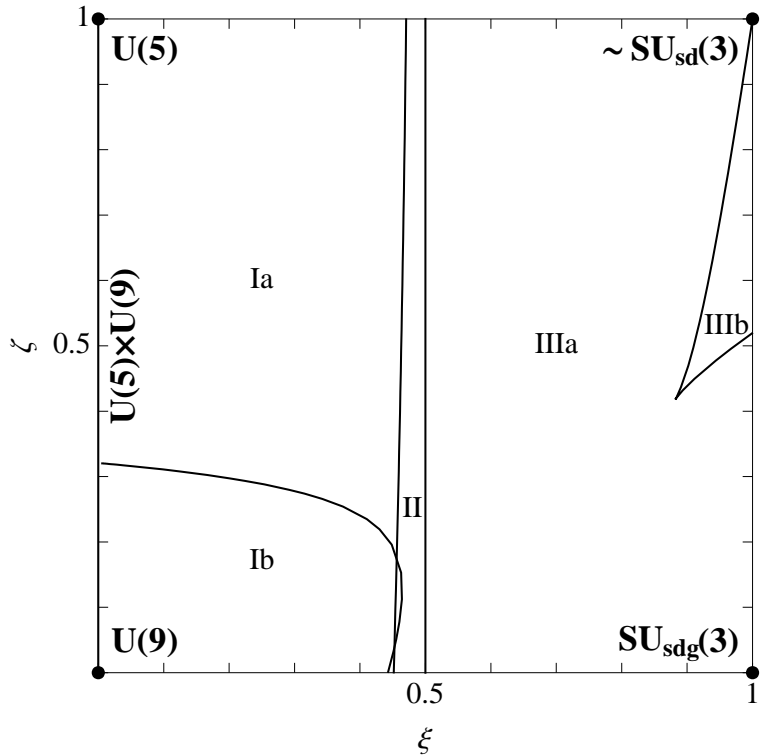}
\caption{The phase diagram in the $(\xi,\zeta)$ plane of the Casten prism for $\chi=-1$.
The lines or curves represent the locus of degenerate critical points with $\gamma_2^*=0$
that separate the different regions as discussed in the text.
Symmetry limits of the \mbox{$sd$-IBM} and \mbox{$sdg$-IBM}
are indicated with black dots.
The entire $\zeta$ axis
corresponds to the ${\rm U}(5)\otimes{\rm U}(9)$ limit of \mbox{$sdg$-IBM}.}
\label{f_u59su3}
\end{figure}
In general, critical points can only be obtained
through a numerical solution of the equations~(\ref{e_crit3}) and~(\ref{e_deg3}),
which, for a given $\chi$, can be achieved as follows
[with reference to the classical limit~(\ref{e_clim})
expressed in units of $\kappa(N-1)$].
Since the hamiltonian is linear in $\epsilon_d$ and $\epsilon_g$,
it is straightforward to obtain as solutions of the first two equations in~(\ref{e_crit3})
analytic expressions $\epsilon_d(\beta_2,\beta_4)$ and $\epsilon_g(\beta_2,\beta_4)$.
Furthermore, the solution of the third equation in~(\ref{e_crit3})
is also known in closed form, see~(\ref{e_solg2}).
As a result, the ``unknowns'' $\epsilon_d$, $\epsilon_g$ and $\gamma_2$
can be eliminated from~(\ref{e_deg3})
to yield a single equation in $\beta_2$ and $\beta_4$.
This equation cannot be solved analytically
but, for a given value of $\beta_2$, and always for fixed given $\chi$,
a numerical solution in $\beta_4$ can be found,
and from there $\gamma_2$, $\epsilon_d$ and $\epsilon_g$ 
are determined.
Transforming from the $(\epsilon_d,\epsilon_g,\kappa(N-1))$
to the $(\xi,\zeta)$ variables,
one thus finds parametric curves in $(\xi,\zeta)$ space
with $\beta_2$ as parameter.
Identical results are found
by inverting the roles of $\beta_2$ and $\beta_4$,
in which case the parametric curves in $(\xi,\zeta)$ space
have $\beta_4$ as parameter.

The solution for $\chi=-1$ already displays a surprising richness
which is illustrated in figure~\ref{f_u59su3}.
The different curves in the figure
represent the locus of degenerate critical points with $\gamma_2^*=0$,
that is, degenerate critical points
that satisfy the first of the equations~(\ref{e_solg2}).
The physical portion of the phase diagram with $0\leq\xi\leq1$ and $0\leq\zeta\leq1$
displays several regions distinguished by qualitatively different potentials.
We may characterize a potential by listing all its critical points
together with the signs of the eigenvalues of the stability matrix
which distinguishes minima from maxima from saddle points.
Hence we introduce for each critical point
a notation $(\lambda_1,\lambda_2,\lambda_3)$ with $\lambda_i=\pm$.
A minimum, for example, is denoted as $(+,+,+)$.
A special situation arises
for spherical critical points with $(\beta_2^*,\beta_4^*)=(0,0)$
which can be characterized by the signs of the eigenvalues $(\lambda_1,\lambda_2)$
of the $2\times2$ stability matrix~(\ref{e_stab2}).
The different regions of figure~\ref{f_u59su3}
then have the following potentials:
\begin{itemize}
\item
(Ia) $(+,+)$, $(+,-,-)$, $(-,-,-)$;
\item
(Ib) $(+,+)$, $(+,+,-)$, $(-,-,-)$;
\item
(II) $(+,+)$, $(+,+,+)$, $(+,+,-)$, $(+,-,-)$, $(-,-,-)$;
\item
(IIIa) $(+,-)$, $(+,+,+)$, $(+,+,-)$, $(+,-,-)$, $(-,-,-)$;
\item
(IIIb) $(+,-)$, $(+,+,+)$, $(+,+,-)$, $(+,-,-)^2$, $(-,-,-)^2$.
\end{itemize}
We emphasize that this is not the complete classification
since it only concerns critical points with $\gamma_2^*=0$;
an even more complex diagram is found
if solutions from the second of the equations~(\ref{e_solg2}) are included.
In the study of phase transitions of quantum systems
we are, however, primarily interested in the evolution
of {\em global and local minima} of the potential.
Therefore, we may for our purpose here
ignore the evolution of saddle points and maxima,
and concentrate on minima.
From this point of view the phase diagram
is divided into three regions characterized by potentials with
(I) a spherical minimum,
(II) a spherical and a deformed minimum,
and (III) a deformed minimum.
Also, if we confine our attention
to the global and local minima of the potential only,
no degenerate critical point with $\gamma_2\neq0$
is found to occur in the physical region of the phase diagram
with $0\leq\xi\leq1$ and $0\leq\zeta\leq1$.
From now on our analysis will be confined
to the partition of the phase diagram
into regions with potentials that differ in the properties of their minima. 

The phase diagram for the ${\rm U}(5)\otimes{\rm U}(9)$--SO(15) transition
of subsection~\ref{ss_u59so15}
is characterized by two main regions (I) and (II)
(see figure~\ref{f_u59so15})
since (IIa) and (IIb) are distinguished
by differences in saddle points and maxima only.
With respect to the result of subsection~\ref{ss_u59so15}
we note that the ${\rm U}(5)\otimes{\rm U}(9)$--SU(3) transition
has an extra region (II) which is characterized
by the coexistence of a spherical and a deformed minimum.
This result is consistent with what is found in the \mbox{$sd$-IBM}
and, moreover, we observe that,
for realistic boson energies with $\epsilon_g>\epsilon_d$,
the actual value of the $g$-boson energy
has little influence on the size of the coexistence region.

\begin{figure}
\centering
\includegraphics[width=15cm]{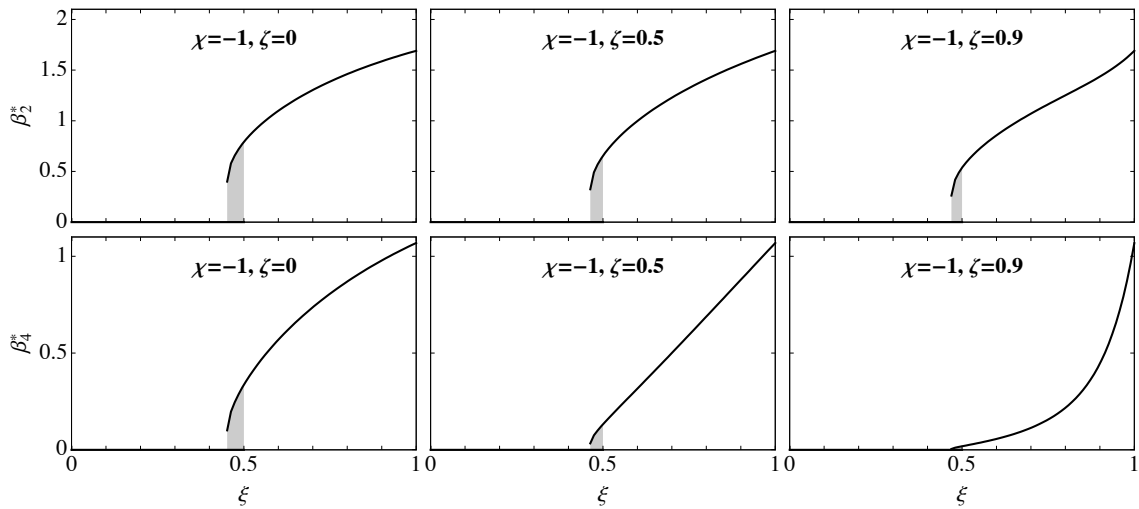}
\caption{
The evolution of the deformation parameters at equilibrium,
$\beta_2^*$ and $\beta_4^*$,
along three different paths in the $(\xi,\zeta)$ plane
with $0\leq\xi\leq1$ and $\zeta=0$, 0.5 and 0.9, respectively.
In each case the region of coexistence
of a spherical and a deformed minimum
is indicated by the grey area.}
\label{f_b2b4}
\end{figure}
Although one does not have analytic expressions
for the deformation parameters at equilibrium,
$\beta_2^*$ and $\beta_4^*$,
these can be obtained numerically
and typical results are shown in figure~\ref{f_b2b4}
for three different paths in the $(\xi,\zeta)$ plane.
The most striking evolution
is that of the behaviour of $\beta_4^*$ as a function of $\zeta$.
It is clear from the figure
that a first-order transition occurs in $\beta_4^*$
but the discontinuity in $\beta_4^*$
becomes smaller for $\zeta\rightarrow1$
(or $\epsilon_g\rightarrow\infty$).

\subsection{The SU$_-$(3)--SO(15)--SU$_+$(3) transition}
\label{ss_so15su3}
The ${\rm SU}_-(3)$--SO(15)--${\rm SU}_+(3)$ transition
is obtained for $\eta=+\infty$ (or $\xi=1$)
while $\chi$ varies from $-1$ to $+1$.
In the limit $\eta\rightarrow+\infty$
there is no dependence of the potential
on either of the boson energies $\epsilon_d$ or $\epsilon_g$,
and hence the search for degenerate critical points
is independent of $\zeta$.
It is easy to show that for $\chi=0$
the equations~(\ref{e_crit3}) are satisfied for ${\beta_2^*}^2+{\beta_4^*}^2=1$
and arbitrary $\gamma_2^*$.
The potential is not only flat in $\gamma_2$
but in addition behaves like a Mexican hat in $(\beta_2,\beta_4)$ around (0,0).
Furthermore, for ${\beta_2^*}^2+{\beta_4^*}^2=1$,
the condition~(\ref{e_deg3}) for a degenerate critical point
is satisfied as well.
Thus we find that all points with $\chi=0$ and arbitrary $\zeta$
are degenerate critical.
On this line the potential is flat in $\gamma_2$
and in the ${\rm SU}_-(3)$--SO(15)--${\rm SU}_+(3)$ transition
it tilts over from a prolate minimum with $\gamma_2^*=0$ for $\chi<0$
to an oblate one with $\gamma_2^*=\pi/3$ for $\chi>0$.
No other degenerate critical points occur
for values of $\chi$ between $-1$ and $+1$.

\section{The complete phase diagram}
\label{s_comp}
We are now in a position to piece together the entire phase diagram
for $0\leq\xi\leq1$, $-1\leq\chi\leq+1$ and $0\leq\zeta\leq1$.
Since the change $\chi\rightarrow-\chi$
is equivalent to the simultaneous change $\beta_4\rightarrow-\beta_4$
and $\gamma_2\rightarrow\pi/3-\gamma_2$,
it is in fact sufficient to consider only positive or negative values of $\chi$:
the phase diagram is mirror-symmetric
with respect to the U(5)--U(9)--SO(15)--SO(6) plane.
In the following we consider the range $-1\leq\chi\leq0$.

An expansion of the potential up to third order,
\begin{eqnarray}
\lefteqn{V(\beta_2,\beta_4,\gamma_2;\eta,\chi,r)
\approx(1-4\eta)\beta_2^2
+[r-4\eta(1-\chi^2)]\beta_4^2}
\nonumber\\&&\qquad\textstyle
+\frac{22}{7}\sqrt{\frac 5 7}\chi\eta\beta_2^3\cos3\gamma_2
+\frac{72}{7}\sqrt{\frac 2 7}\chi\eta\beta_2^2\beta_4
+\frac{20}{7}\sqrt{\frac 5 7}\chi\eta\beta_2\beta_4^2\cos3\gamma_2,
\label{e_exp}
\end{eqnarray}
or, alternatively, the expression for the stability matrix at $\beta_2^*=\beta_4^*=0$,
\begin{equation}
{\cal H}_2(0,0;\xi,\chi=0,\zeta)=
\left[\begin{array}{cc}
\displaystyle
\frac{2(1-2\xi)}{1-\xi}&0\\[2ex]
0&\displaystyle\frac{2\{1-\xi[2-\zeta-\chi^2(1-\zeta)]\}}{(1-\xi)(1-\zeta)}
\end{array}\right],
\label{e_stab20}
\end{equation}
shows that at the spherical point the potential has
(I) a minimum for $\xi<1/2$ (or $\eta<1/4$),
(IIa) a saddle point for $1/2<\xi<1/[2-\zeta-\chi^2(1-\zeta)]$
[or $1/4<\eta<r/(4-4\chi^2)$]
and (IIb) a maximum for $\xi>1/[2-\zeta-\chi^2(1-\zeta)]$ [or $\eta>r/(4-4\chi^2)$],
always at $\beta_2^*=\beta_4^*=0$.
The situation is similar to the transitions discussed in subsection~\ref{ss_u59so15},
with the only difference that now the separation line between the regions (IIa) and (IIb)
depends on the value of $\chi$
while for the ${\rm U}(5)\otimes{\rm U}(9)$--SO(15) transition
we have $\chi=0$.
If we confine ourselves to an analysis
of the properties of the {\em minima} of the potential,
we may ignore the separation between (IIa) and (IIb)
and treat them as a single region.

A locus of points of particular interest is defined by the condition
that {\em two} eigenvalues of the stability matrix vanish.
This is possible if there are at least two order parameters
and at least three control parameters, as is the case presently,
and leads to a catastrophe function of the type $D_{\pm4}$~\cite{Gilmore81}.
From the stability matrix~(\ref{e_stab20}) it is seen that this happens
at $\beta_2^*=\beta_4^*=0$ for $\xi=1/2$ (or $\eta=1/4$)
and $\chi^2=\zeta/(\zeta-1)$
(or $\chi^2=(\epsilon_d-\epsilon_g)/\epsilon_d$).
The locus of $D_{\pm4}$ catastrophes is represented by a curve
in the $(\xi,\chi,\zeta)$ space
but it is entirely situated in the non-physical portion of this space
(for $\epsilon_g\leq\epsilon_d$, {\em below} the Casten prism of figure~\ref{f_prism})
with the exception of the single point $\xi=1/2$, $\chi=\zeta=0$
which is half-way between U(9) and SO(15).

\begin{figure}
\centering
\includegraphics[width=10cm]{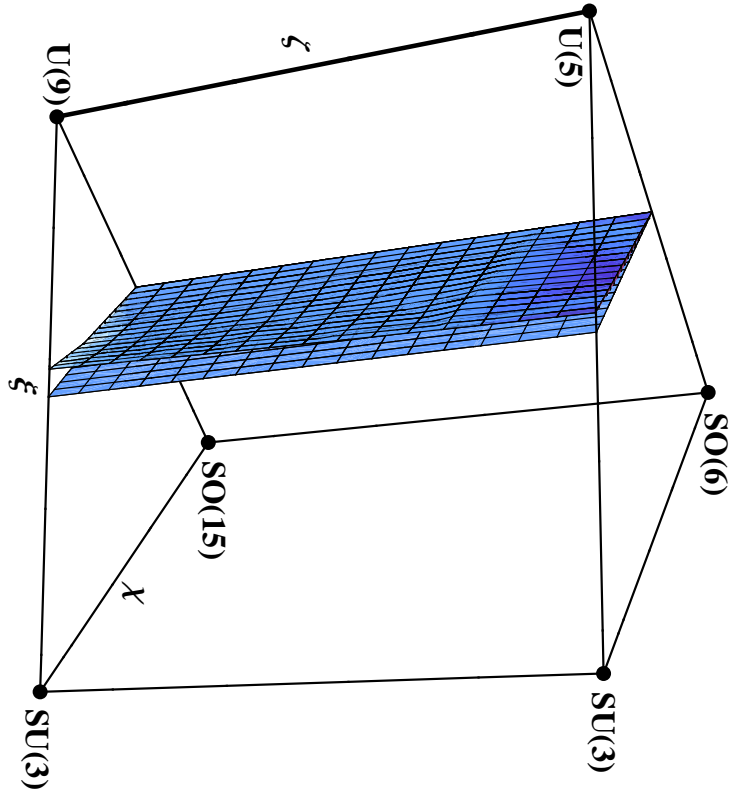}
\caption{Phase diagram of the Casten prism associated
with the hamiltonian~(\ref{e_cassdg}) of the \mbox{$sdg$-IBM}.
The two planes represent the locus of degenerate critical points
that separate the spherical and deformed regions,
and enclose the region of coexistence.
The different symmetry limits
of the \mbox{$sd$-IBM} and \mbox{$sdg$-IBM}
are indicated with black dots.}
\label{f_phase}
\end{figure}
We therefore come to the conclusion
that nothing spectacular happens within the Casten prism
which, when it comes to the properties of {\em minima},
is divided by two surfaces in to three regions (see figure~\ref{f_phase}).
The first surface is a plane
obtained for $\xi=1/2$, $-1\leq\chi\leq0$ and $0\leq\zeta\leq1$.
The second is a parametric surface
with parameters $(\beta_2,\chi)$ [or $(\beta_4,\chi)$],
constructed with the technique explained in subsection~\ref{ss_u59su3}.
This situation closely parallels what is obtained in \mbox{$sd$-IBM}
and, moreover, it is seen from figure~\ref{f_phase}
that the size of the coexistence region is not greatly influenced
by the ratio of boson energies $\epsilon_g/\epsilon_d$ (or $\zeta$).
We want to emphasize, however, that this is not a trivial conclusion
but that this result depends on our choice of hamiltonian
and of the domain available to the parameters $\xi$, $\chi$ and $\zeta$
which are guided by physics arguments.
If we were to allow, for example, boson energies with $\epsilon_g<\epsilon_d$,
a much more complex phase diagram would be uncovered
which strongly departs from the one found in \mbox{$sd$-IBM}.

\section{Conclusion}
\label{s_conc}
We summarize the three main assumptions
at the basis of our geometric analysis of the \mbox{$sdg$-IBM}.
\begin{enumerate}
\item
The general parametrization of a shape
with quadrupole and hexadecapole deformation
is reduced to one in terms of three variables $(\beta_2,\beta_4,\gamma_2)$.
\item
A simplified $sdg$ hamiltonian is considered
which is intermediate between four dynamical-symmetry limits of U(15),
namely ${\rm U}(5)\otimes{\rm U}(9)$, ${\rm SU}_\pm(3)$  and SO(15).
\item
Parameters of this simplified $sdg$ hamiltonian
are restricted to physically acceptable values.
\end{enumerate}
Under the above assumptions we find 
that the phase diagram is divided into three regions
characterized by potentials with
(I) a spherical minimum,
(II) a spherical and a deformed minimum,
and (III) a deformed minimum.
The deformed minima are either prolate or oblate
depending on the sign of the parameter $\chi$ in the quadrupole operator.
No transition towards a stable triaxial shape is found.

\section*{Acknowledgments}
This work has been carried out in the framework of CNRS/DEF project N 19848.
S~Z thanks the Algerian Ministry of High Education and Scientific Research for financial support.


\section{References}

\begin{thebibliography}{99}
\bibitem{Gilmore79}
Gilmore~R 1979
{\it J.\ Math.\ Phys.\ } {\bf20} 891

\bibitem{Arima75}
Arima~A and Iachello~F 1975 
{\it Phys.\ Rev.\ Lett.\ } {\bf35} 1069

\bibitem{Arima76}
Arima~A and Iachello~F 1976
{\it Ann.\ Phys.\ (NY)} {\bf99} 253

\bibitem{Arima78}
Arima~A and Iachello~F 1978
{\it Ann.\ Phys.\ (NY)} {\bf111} 201

\bibitem{Arima79}
Arima~A and Iachello~F 1979
{\it Ann.\ Phys.\ (NY)} {\bf123} 468

\bibitem{Iachello87}
Iachello~F and Arima~A 1987
{\it The Interacting Boson Model}
(Cambridge: Cambridge University Press)

\bibitem{Dieperink80}
Dieperink~A~E~L, Scholten~O and Iachello~F 1980
{\it Phys.\ Rev.\ Lett.\ } {\bf44} 1747

\bibitem{Ginocchio80}
Ginocchio~J~N and Kirson~M~W 1980
{\it Phys.\ Rev.\ Lett.\ } {\bf44} 1744

\bibitem{Bohr80}
Bohr~A and Mottelson~B~R 1980
{\it Phys.\ Scripta} {\bf22} 468

\bibitem{BM75}
Bohr~A and Mottelson~B~R 1975
{\it Nuclear Structure. II Nuclear Deformations}
(Reading, Massa\-chusetts: Benjamin)

\bibitem{Feng81}
Feng~D~H, Gilmore~R and S.R. Deans~S~R 1981
{\it Phys.\ Rev.\ C} {\bf23} 1254

\bibitem{Lopez96}
L\'opez-Moreno~E and Casta\~nos~O 1996
{\it Phys.\ Rev.\ C} {\bf54} 2374

\bibitem{Iachello98}
Iachello~F, Zamfir~N~V and Casten~R~F 1998
{\it Phys.\ Rev.\ Lett.\ } {\bf81} 1191

\bibitem{Casten99}
Casten~R~F, Kusnezov~D and Zamfir~N~V 1999
{\it Phys.\ Rev.\ Lett.\ } {\bf82} 5000

\bibitem{Cejnar00}
Cejnar~P and Jolie~J 2000
{\it Phys.\ Rev.\ E} {\bf61} 6237

\bibitem{Iachello00}
Iachello~F 2000
{\it Phys.\ Rev.\ Lett.\ } {\bf85} 3580

\bibitem{Iachello01}
Iachello~F 2001
{\it Phys.\ Rev.\ Lett.\ } {\bf87} 052502

\bibitem{Casten00}
Casten~R~F and Zamfir~N~V 2000
{\it Phys.\ Rev.\ Lett.\ } {\bf85} 3584

\bibitem{Casten01}
Casten~R~F and Zamfir~N~V 2001
{\it Phys.\ Rev.\ Lett.\ } {\bf87} 052503

\bibitem{Frank02}
Frank~A, Alonso~C~E and Arias~J~M, 2002
{\it Phys.\ Rev.\ C} {\bf65} 014301

\bibitem{Arias03}
Arias~J~M, Alonso~C~E, Vitturi~A, Garc\'\i a-Ramos~J~E, Dukelsky~J and Frank~A 2003
{\it Phys.\ Rev.\ C} {\bf68} 041302(R)

\bibitem{Garcia03}
Garc\'\i a-Ramos~J~E, Arias~J~M, Barea~J and Frank~A 2003
{\it Phys.\ Rev.\ C} {\bf68} 024307

\bibitem{Leviatan03}
Leviatan~A and Ginocchio~J~N 2003
{\it Phys.\ Rev.\ Lett.\ } {\bf90} 212501

\bibitem{Garcia05}
Garc\'\i a-Ramos~J~E, Dukelsky~J and Arias~J~M 2005
{\it Phys.\ Rev.\ C} {\bf72} 037301

\bibitem{Jolie01}
Jolie~J, Casten~R~F, von~Brentano~P and Werner~V 2001
{\it Phys.\ Rev.\ Lett.\ } {\bf78} 162501

\bibitem{Jolie03}
Jolie~J and Linnemann~A 2003
{\it Phys.\ Rev.\ C} {\bf68} 031301(R)

\bibitem{Jolie02}
Jolie~J, Cejnar~P, Casten~R~F, Heinze~S, Linnemann~A and Werner~V 2002
{\it Phys.\ Rev.\ Lett.\ } {\bf89} 182502

\bibitem{Cejnar03b}
Cejnar~P, Heinze~S and Jolie~J 2003
{\it Phys.\ Rev.\ C} {\bf68} 034326

\bibitem{Iachello04}
Iachello~F and Zamfir~N~V 2004
{\it Phys.\ Rev.\ Lett.\ } {\bf92} 212501

\bibitem{Cejnar02}
Cejnar~P 2002
{\it Phys.\ Rev.\ C} {\bf65} 044312

\bibitem{Cejnar03}
Cejnar~P 2003
{\it Phys.\ Rev.\ Lett.\ } {\bf90} 112501

\bibitem{Cejnar04}
Cejnar~P and Jolie~J 2004
{\it Phys.\ Rev.\ C} {\bf69} 011301(R)

\bibitem{Heyde04}
Heyde~K, Jolie~J, Fossion~R, De~Baerdemacker~S and Hellemans~V 2004
{\it Phys.\ Rev.\ C} {\bf69} 054304

\bibitem{Rowe04}
Rowe~D~J, Turner~P~S and Rosensteel~G 2004
{\it Phys.\ Rev.\ Lett.\ } {\bf93} 212502

\bibitem{Rowe04b}
Rowe~D~J 2004
{\it Phys.\ Rev.\ Lett.\ } {\bf93} 122502

\bibitem{Rowe04c}
Rowe~D~J 2004
{\it Nucl.\ Phys.\ A} {\bf745} 47

\bibitem{Turner05}
Turner~P~S and Rowe~D~J 2005
{\it Nucl.\ Phys.\ A} {\bf756} 333

\bibitem{Rosensteel05}
Rosensteel~G and Rowe~D~J 2005
{\it Nucl.\ Phys.\ A} {\bf759} 92

\bibitem{Caprio08}
Caprio~M~A, Cejnar~P and Iachello~F 2008
{\it Ann.\ Phys.\ (NY)} {\bf323} 1106

\bibitem{Caprio04}
Caprio~M~A and Iachello~F 2004
{\it Phys.\ Rev.\ Lett.\ } {\bf93} 242502

\bibitem{Frank06}
Frank~A, Van~Isacker~P and Iachello~F 2006
{\it Phys.\ Rev.\ C} {\bf73} 061302(R)

\bibitem{Devi92}
Devi~Y~D and Kota~V~K~B 1992
{\it Pramana J.\ Phys.\ } {\bf39} 413

\bibitem{Devi90}
Devi~Y~D and Kota~V~K~B 1990
{\it Z.\ Phys.\ A} {\bf337} 15

\bibitem{Rohozinski81}
Rohozi\'nski~S~G and Sobiczewski~A 1981
{\it Acta Phys.\ Pol.\ B} {\bf12} 1001

\bibitem{Nazarewicz81}
Nazarewicz~W and Rozmej~P 1981
{\it Nucl.\ Phys.\ A} {\bf369} 396

\bibitem{Isacker81}
Van~Isacker~P and Chen~J-Q 1981
{\it Phys.\ Rev.\ C} {\bf24} 684

\bibitem{Warner82}
Warner~D~D and Casten~R~F 1982
{\it Phys.\ Rev.\ Lett.\ } {\bf48} 1385

\bibitem{Lipas85}
Lipas~P~O, Toivonen~P and Warner~D~D 1985
{\it Phys.\ Lett.\ B} {\bf155} 295

\bibitem{Casten81}
Casten~R~F 1981
{\it Interacting Bose-Fermi Systems in Nuclei},
edited by Iachello~F (New York: Plenum) p~3

\bibitem{Demeyer86}
Demeyer~H, Van der Jeugt~J, Vanden Berghe~G and Kota~V~K~B 1986
{\it J.\ Phys.\ A} {\bf19} L565

\bibitem{Kota87}
Kota~V~K~B, Van der Jeugt~J, Demeyer~H and Vanden Berghe~G 1987
{\it J.\ Math.\ Phys.\ } {\bf28} 1644

\bibitem{Bouldjedri05}
Bouldjedri~A, Van~Isacker~P and Zerguine~S 2005
{\it J.\ Phys.\ G} {\bf31} 1329

\bibitem{Gilmore81}
Gilmore~R 1981
{\it Catastrophe Theory for Scientists and Engineers}
(New York: Wiley)

\end{thebibliography}
\end{document}